\documentclass[submission]{eptcs}
\providecommand{\event}{SOS 2007} 
\usepackage{breakurl}

\usepackage[pdftex]{graphicx}

\usepackage[]{hyperref}
\hypersetup{
  colorlinks=true, linkcolor=black, citecolor=black, urlcolor=black,
}

\title{A Solution to the Flowgraphs Case Study using Triple Graph Grammars and eMoflon}
\author{Anthony Anjorin\thanks{Supported by the `Excellence Initiative' of the German Federal and State Governments and the Graduate School of Computational Engineering at TU Darmstadt.} 
\institute{ 
Technische Universit\"at Darmstadt\\
Real-Time Systems Lab\\ 
Merckstr. 25\\
64283 Darmstadt, Germany\\
\email{anjorin@es.tu-darmstadt.de} 
}  
\and 
Marius Lauder$^*$
\institute{ 
Technische Universit\"at Darmstadt\\ 
Real-Time Systems Lab\\ 
Merckstr. 25\\
64283 Darmstadt, Germany\\
\email{lauder@es.tu-darmstadt.de} 
}}   
\def\titlerunning{A Solution to the Flowgraphs Case Study using TGGs and eMoflon}
\def\authorrunning{A. Anjorin \& M. Lauder}

\begin{document}
\maketitle 

\begin{abstract}
After 20 years of Triple Graph Grammars (TGGs) and numerous actively maintained implementations, there is now a need for challenging examples and success stories to show that TGGs can be used for real-world bidirectional model transformations. 
Our primary goal in recent years has been to increase the expressiveness of TGGs by providing a set of pragmatic features that allow a controlled fallback to programmed graph transformations and Java.  

Based on the \emph{Flowgraphs} case study~\cite{Horn13} of the Transformation Tool Contest (TTC 2013), we present (i) \emph{attribute constraints} used to express complex bidirectional attribute manipulation, (ii) \emph{binding expressions} for specifying arbitrary context relationships, and (iii) \emph{post-processing methods} as a black box extension for TGG rules.
In each case, we discuss the enabled trade-off between guaranteed formal properties and expressiveness.

Our solution, implemented with eMoflon (\url{www.emoflon.org}) our metamodelling and model transformation tool, is available as a virtual machine hosted on Share~\cite{TTC13_eMoflonvdi}.
\end{abstract}

\section{Introduction and Motivation}
\label{sect:intro-and-motivation}

\emph{Triple Graph Grammars} (TGGs)~\cite{Schurr1994} are a declarative, rule-based, bidirectional model transformation language and can be used to specify a \emph{consistency relation} over source and target models, with which various \emph{operational scenarios}, such as a forward and backward transformation, can be automatically supported.
As TGGs have been in use for about 20 years and numerous formal results based on algebraic graph transformation~\cite{Klar2010} have been established, it is now time to tackle convincingly realistic real-world case studies with TGGs.

In our opinion, based on practical experience with our TGG implementation in eMoflon~\cite{Anjorin2011}, what is necessary is to increase the expressiveness of TGGs by providing a pragmatic set of features that allow a controlled \emph{fallback} to a general purpose (transformation) language such as standard Java or programmed graph transformations via, e.g., \emph{Story Driven Modelling} (SDM)~\cite{Fischer2000}. 

Our contribution in this paper is to present a set of advanced TGG features that enable a controlled integration of SDM and Java code in TGG rules.
These features are: (i) \emph{attribute constraints} for complex bidirectional attribute manipulation in TGG rules, (ii) \emph{binding expressions} for expressing complex, possibly recursive context relationships, and (iii) \emph{post-processing methods} that allow for arbitrary post-processing after a TGG rule has been applied.
Using the case-study \emph{Flowgraphs} of the TTC 2013~\cite{Horn13}, we present each feature, explaining when it is necessary and discussing the involved trade-off between expressiveness and formal properties.

\section{Solution With Triple Graph Grammars}
\label{sect:solution-with-tggs}

The \emph{Flowgraphs} case study of the TTC 2013 is a text-to-model transformation involving the static analysis of a small Java subset.
The task is to transform a simple Abstract Syntax Tree~(AST) for Java to a \emph{flowgraph}, i.e., a model containing explicit information about the control and data flow in the program.  
In this context, \emph{bidirectionality} is a crucial requirement to reflect program manipulations (refactorings, quick-fixes), applied via model transformations, back in the code.
To show that eMoflon supports using TGGs directly with standard parser and unparser technology (including XML), we modify the case study slightly by using ANTLR~\cite{Parr2007} and StringTemplate~\cite{Parr2004} for parsing and unparsing Java code, respectively.
Our TGG rules operate directly on the AST produced by ANTLR without a model parser such as EMFText~\cite{Heidenreich2009}.
 
\begin{figure}[!hb]
\begin{center} 
  \includegraphics[width=0.92\textwidth]{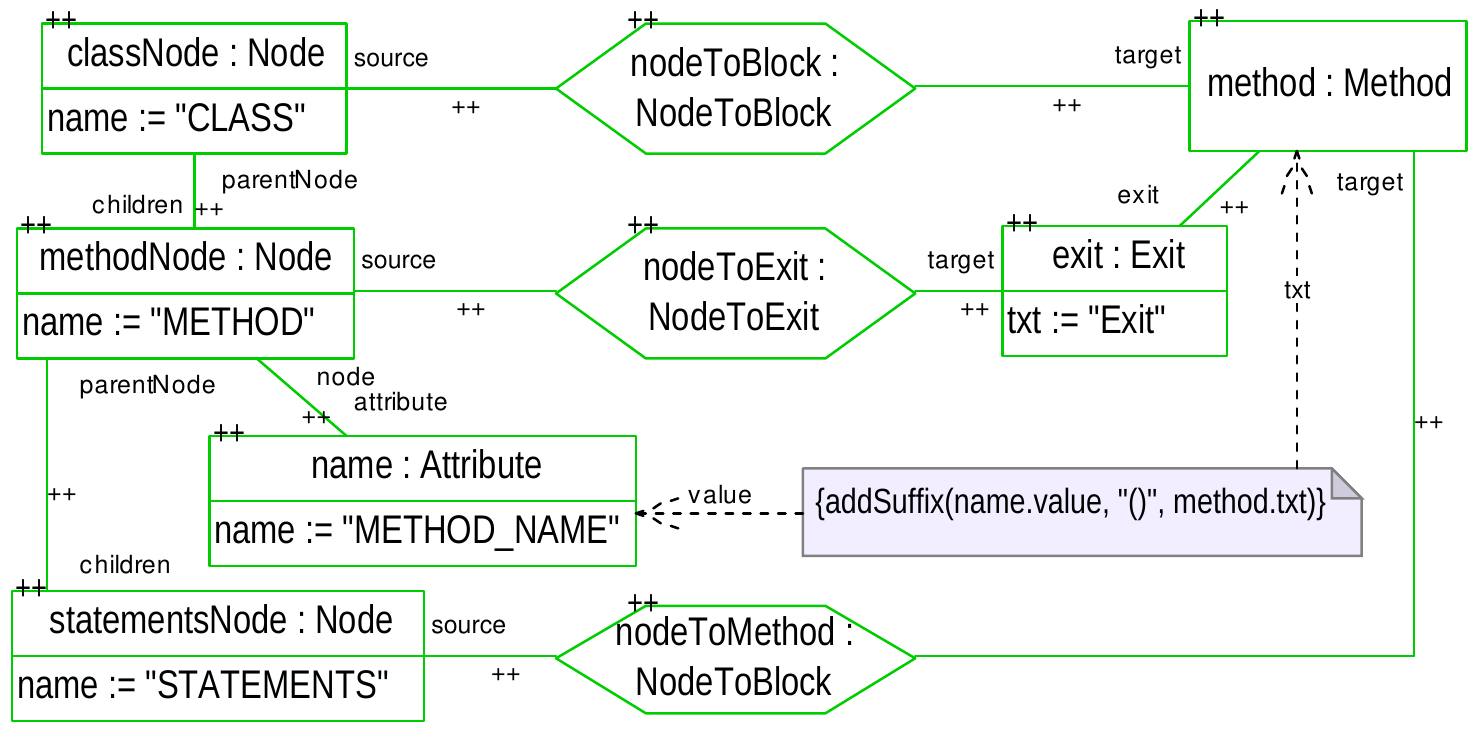}
  \caption{TGG axiom \texttt{MethodRule} for creating the corresponding tree structure for \texttt{Methods}}
  \label{fig:axiom}
\end{center}
\end{figure}

A TGG consists of a set of \emph{rules} describing how source and target models evolve simultaneously.
Each \emph{TGG rule} consists of \emph{elements} (objects and links), typed according to the classes and references in the respective metamodels. 
The TGG rule \textsf{MethodRule} (Fig.~\ref{fig:axiom}) creates the root element of the AST representing a class containing a single method (on the left hand side), and a \textsf{Method} with an explicit \textsf{Exit} element in the model (on the right hand side).
As can be seen from the rule, the AST metamodel (not shown explicitly) consists basically of labelled nodes with children and attributes.

\textsf{MethodRule} is referred to as an axiom as it only creates elements (green with a \textsf{``++''} markup) and does not require any context (precondition).
Note how \emph{correspondence} elements such as \textsf{nodeToBlock}, depicted visually as hexagons, are created to connect AST elements (\textsf{classNode}) with model elements (\textsf{method}).
These elements are used in other rules as traceability links and conform to a simple correspondence metamodel, connecting source and target metamodels.
The triple of source, correspondence, and target metamodels is referred to as a \emph{TGG schema}. 

Simple attribute manipulation can be expressed as \emph{attribute assignments} such as \textsf{txt = ``Exit''} in the \textsf{exit} element.
Note that this is automatically interpreted as an assertion for the backward transformation, where the model is parsed and the AST is created.

More complex attribute manipulation can be specified using bidirectional \emph{attribute constraints}, such as the constraint \textsf{addSuffix(name.value, ``()'', method.txt)} in \textsf{MethodRule}, which expresses that the value of the \textsf{name} attribute in the AST must be equal to the \textsf{txt} attribute of the \textsf{method} element in the model, after adding ``()'' as a suffix.
Attribute constraints such as \textsf{addSuffix} are interpreted appropriately when compiling the TGG rule for each operational scenario (e.g., forward, backward) and can thus be expressed in a direction agnostic manner fitting to the rest of the TGG rule. 

\begin{figure}[!ht]
\begin{center}
  \includegraphics[width=0.86\textwidth]{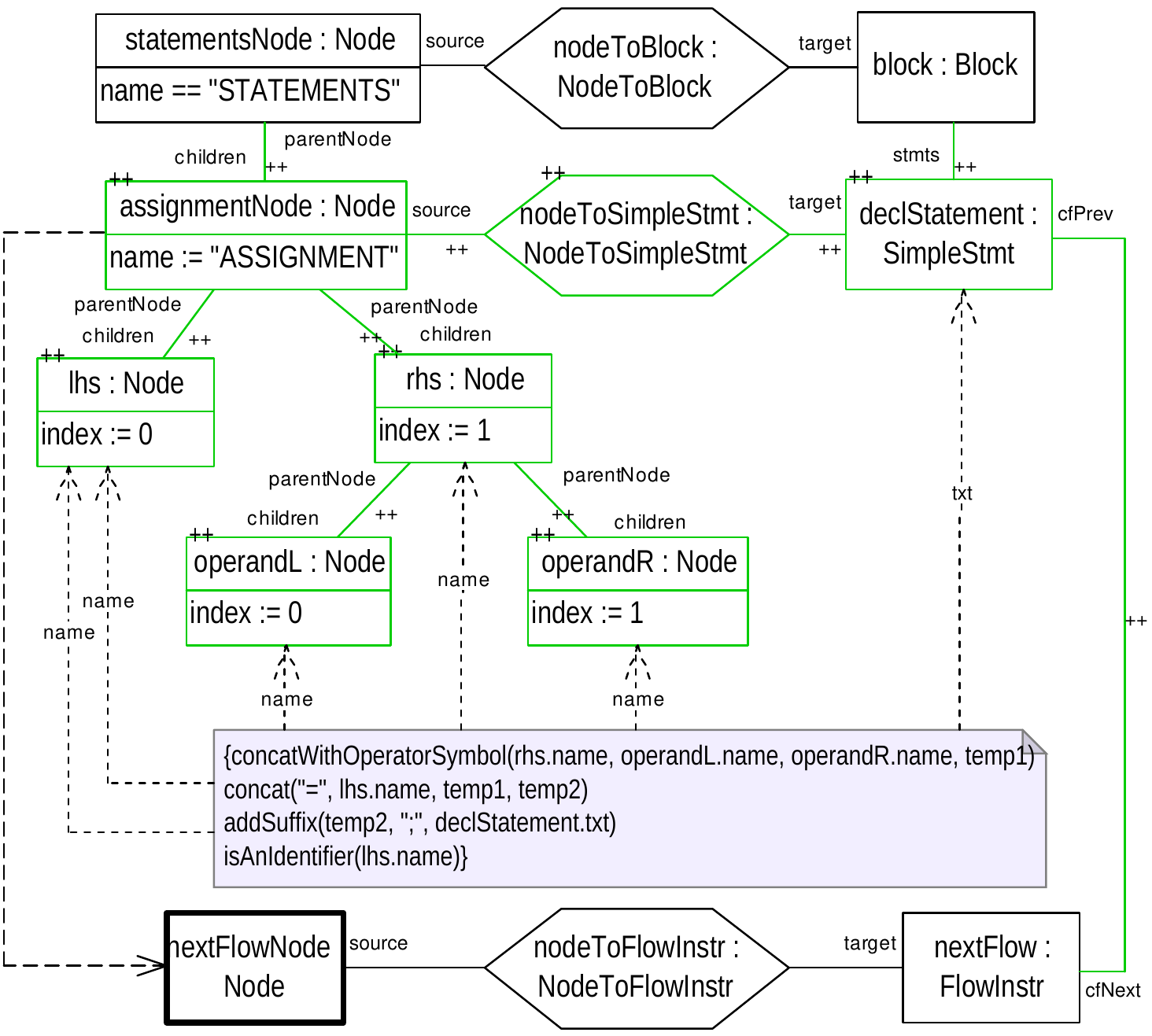}
  \caption{TGG rule \texttt{AssignmentWithExpRule} for creating assignment statements}
  \label{fig:rule}
\end{center}
\end{figure}

A second TGG rule \textsf{AssignmentWithExpRule} (Fig.~\ref{fig:rule}) handles, for example, statements of the form \textsf{``a = b + 3;''} represented in the AST as an assignment tree with \textsf{lhs = ``a''}, \textsf{rhs = ``+''}, \textsf{operandL = ``b''}, and \textsf{operandR = ``3''}.
To test the flexibility of the transformation language, the case study requires an embedded \emph{model-to-text} transformation to transform the tree structure back to text and store it as the attribute value of the \textsf{SimpleStmt} element in the model.
This is accomplished in \textsf{AssignmentWithExpRule} with a set of attribute constraints, referred to as the \emph{attribute Constraint Satisfaction Problem (CSP)} of the TGG rule.
How this works is best explained using a concrete example;
To handle \textsf{``a = b + 3;''} in the forward transformation, the set of constraints is sorted and solved as follows:

\begin{verbatim}
	isAnIdentifier("a") => true
	concatWithOperatorSymbol("+", "b", "3", temp1) => temp1 = "b + 3"
	concat("=", "a", "b + 3", temp2) => temp2 = "a = b + 3"
	addSuffix("a = b + 3", ";", declStm.txt) => declStm.txt = "a = b + 3;"
\end{verbatim}

The constraints \textsf{isAnIdentifier} and \textsf{concatWithOperatorSymbol} are \emph{user defined constraints} and were implemented in Java specifically for the Flowgraphs case study.
All other constraints are \emph{library constraints} and are directly available for use in our tool.
In this way, the set of constraints can be seamlessly extended by the user, reused, and combined with other constraints in different TGG rules.
The constraint \textsf{isAnIdentifier} ensures that the rule is not applied for statements such as \textsf{``int a = b + 3''}, where \textsf{``int a''} is not an identifier and is handled with a different rule for declarations.
The second user defined constraint \textsf{concatWithOperatorSymbol} is an extension of the normal \textsf{concat} constraint and is able to split expressions using a list of supported operators as potential characters for splitting/concatenation as required.
The attribute CSP is sorted and solved differently but analogously for the backward transformation.

Formally, attribute CSPs serve as \emph{application conditions} for TGG rules, and, as access is restricted to only attribute values, most formal results for TGGs still hold as we have shown in previous work~\cite{Anjorin2012b}.
Attribute constraints are implemented in Java and thus allow a controlled integration of Java code for complex attribute manipulation in TGG rules.
 
Note how the rule requires context elements (black without any markup) that must already exist and have been created by applying a different rule such as \textsf{MethodRule}.
In the model, \textsf{AssignmentWithExpRule} requires (i) the parent \textsf{block} element that is to contain the created assignment statement \textsf{declStatement}, and (ii) the succeeding element \textsf{nextFlow} in the flowgraph.

Although the created \textsf{cfPrev} and \textsf{cfNext} references are explicit in the \emph{FlowGraph} metamodel, determining the corresponding next node \textsf{nextFlowNode} in the AST structure is more challenging and requires a non-trivial recursive search in the AST.
In our approach, this search is specified using SDM or Java, and is integrated in the TGG rule via a virtual link referred to as a \emph{binding expression}.
In \textsf{AssignmentWithExpRule}, the binding expression is the link from \textsf{assignmentNode} to \textsf{nextFlowNode}, which is visualized as a dashed arrow leading to an object with a bold border. 
Using binding expressions, complex context relationships in TGG rules can be indicated by requiring a virtual link that does not really exist according to the corresponding metamodel.  
Our tool automatically creates a helper method, which is invoked when navigating along this link, and can be implemented with SDM or Java.

Formally, binding expressions are equivalent to a separate pre-processing step in which all virtual links are created explicitly in an instance of an appropriately augmented metamodel.  
The TGG rules are then viewed as operating on this augmented metamodel. 
In practice, however, it is much more convenient to ``find'' these links \emph{on-the-fly} as required.
In this manner, binding expressions allow the controlled integration of SDM or Java code in TGG rules to determine complex context relationships.
%

As a final feature, every TGG rule can invoke a \emph{post-processing method} (implemented with SDM or Java) to perform some final tasks that are difficult or impossible to specify directly in the TGG rule.
In \textsf{AssignmentWithExpRule}, the index of the created \textsf{assignmentNode} has to be set correctly for the backward transformation, which requires a non-trivial recursive search in the tree structure due to if/else, loop, and return/break statements.

Post-processing methods are clearly a black-box extension and beyond any formal reasoning.
In our practical experience, however, post-processing can be kept in most cases to a bare minimum by using a combination of attribute constraints and binding expressions. 
\section{Related Work}
\label{sect:related-work}
 
Related approaches can be divided into three main groups: (i) Other ways of increasing the expressiveness of TGGs, (ii) Other bidirectional approaches, and (iii) Achieving bidirectionality with unidirectional transformation languages.   

\vspace{0.2cm} 
\noindent \textbf{(i) TGGs with a separated pre- and post-processing step:}  
A justified question is if it is not better to use TGGs with a clearly separated pre-processing and post-processing step instead of our features for integrating auxiliary methods in the TGG rules.
Our approach, however, avoids an extra traversal through the input and output models as \emph{binding expressions} can be used as required to determine context relationships before a TGG rule is applied and a \emph{post-processing method} to complete the rule after application.
In this way, the traversal strategy applied by the TGG control algorithm can be used for \emph{on-the-fly} processing as required.
Our practical experience is that users tend to regard TGGs as not being worth the effort if complex pre-processing and post-processing with a separate traversal is necessary.
With our approach, the TGG is more in focus and is used to structure the transformation in an iterative manner.
  
\vspace{0.1cm}
\noindent \textbf{(ii) Other bidirectional languages and tools:}
For a detailed survey of bidirectional languages other than TGGs we refer to \cite{Stevens2008a}.
Compared to other bidirectional approaches, TGGs are advantageous as there exist multiple, actively maintained TGG implementations (interpretative, generative and hybrid approaches) with different strengths and weaknesses.
Furthermore, most TGG implementations are Ecore/EMF/Eclipse compatible and are thus suitable in an MDE context together with other EMF tools and infrastructure.
In a different context, however, other tools are probably more suitable, e.g., for bidirectional XML manipulation \cite{Nentwich2002} and for bidirectional string and tree manipulation \cite{Bohannon2008}.
Depending on the application scenario, it might be more appropriate to derive a backward transformation from a given forward transformation~\cite{Hidaka2011} as opposed to describing the simultaneous build-up of model triples. 

\vspace{0.1cm}
\noindent \textbf{(iii) Combination of unidirectional transformation languages:}  
  A combination of unidirectional transformation languages as an alternative to a bidirectional language is advantageous for obvious reasons: Unidirectional transformation languages are typically better established, stable, expressive and have better tool support.
  We, however, regard transformation languages with explicit support for bidirectionality as superior as they enable a high-level specification.  
In the case of TGGs, the specified consistency relation can be used to automatically derive other useful \emph{operational} transformations to support conformance testing, consistency checking and link creation for existing source and target models, and incremental model synchronization~\cite{Lauder2012}.
Supporting these scenarios and guaranteeing suitable formal properties~\cite{Klar2010} is quite challenging using separate forward and backward transformations.   
\section{Future Work}
\label{sect:future-work}


In the near future we will establish, compare, and experiment with various ideas for modularity concepts to improve the maintainability of TGG rules.
Efficiency and scalability of transformations with TGGs are also crucial points that must be improved.
We are currently using TGGs for an increasing number of non-trivial transformations including industrial projects together with Siemens AG and internally (i.e., as part of eMoflon) for a bootstrap of a textual syntax for Ecore/SDM/TGGs with TGGs.
We aim to establish a transformation zoo for TGGs consisting of numerous representative examples.
\clearpage 

\bibliographystyle{eptcs}
\bibliography{lib,lib-manual}

\end{document}